\documentstyle[12pt,aasms4,epsfig]{article}
\newcommand{\be}{\begin{equation}}
\newcommand{\ee}{\end{equation}}
\newcommand{\bea}{\begin{eqnarray}}
\newcommand{\eea}{\end{eqnarray}}
\def\lt{\left}
\def\rt{\right}

\def\gte{\,\lower.6ex\hbox{$\buildrel >\over \sim$} \, }
\def\lte{\,\lower.6ex\hbox{$\buildrel <\over \sim$} \, }

\received{}
\accepted{}
\journalid{}{}
\articleid{}{}

\lefthead{M.P. D\c{a}browski and F.E. Schunck}
\righthead{Boson Stars as Gravitational Lenses}

\begin{document}

\title{Boson Stars as Gravitational Lenses}

\author{Mariusz P.~D\c{a}browski$^{1,2}$ and Franz E. Schunck$^{1}$}

\affil{$^1$Astronomy Centre, University of Sussex, Falmer,
Brighton BN1 9QJ, UK\\
$^2$Institute of Physics, University of Szczecin, Wielkopolska 15,
70-451 Szczecin, Poland}

%\author{Franz E. Schunck$^{1}$}
%\affil{$^1$Astronomy Centre, University of Sussex, Falmer,
%Brighton BN1 9QJ, UK}

\begin{abstract}

We discuss boson stars as possible gravitational lenses and study the
lensing effect by these objects made of scalar particles. The mass and
the size of a boson star may vary from an individual Newtonian
object similar to the Sun to the general relativistic size and
mass of a galaxy close to its Schwarzschild radius. We assume boson
stars to be transparent which allows the light to pass through them
though the light is gravitationally deflected.

We assume boson stars of the mass $M = 10^{10}M_\odot$ to be on
non-cosmological distance from the observer. We discuss the lens
equation for these stars as well as the details of magnification.
We find that there are typically three images of a star but the
deflection angles may vary from arcseconds to even degrees.
There is one tangential critical curve (Einstein ring) and one radial
critical curve for tangential and radial magnification, respectively.
Moreover, the deflection angles for the light passing in the gravitational
field of boson stars can be very large (even of the order of degrees)
which reflects the fact they are very strong relativistic objects.
We derive a suitable formula for the lens equation for such
large deflection angles. Although the large deflection angle images are
highly demagnified, their existence in the area of the tangential critical
curve may help in observational detection of suitable lenses possessing
characteristic features of boson stars which
could also serve as a direct evidence for scalar fields in the universe.

\end{abstract}

\keywords{gravitational lensing --- gravitation --- stars: general}

\section{Introduction}

Boson stars are astrophysical objects which are composed of scalar particles
formally described by a complex scalar field coupled minimally to gravity
(Mielke \& Schunck 1998). Real scalar fields became extremely important once
the mutual connections between cosmology and particle physics were uncovered.
The first attempt to admit a scalar field to gravity was made by
Brans \& Dicke (1961) in order to vary the gravitational force with time.
The most intensive application of a scalar field is to the early universe
scenarios. This is first of all to inflation (e.g.~Linde 1990) driven
by the potential energy of the real scalar field (inflaton) and to
phase transitions  together with the production of topological defects
(e.g.~Vilenkin \& Shellard 1994). It is also applied in the theories
of superstrings where it is called the
dilaton (changes strength of gravity) and leads to the so-called
superinflation driven by the kinetic energy of the scalar field
(see e.g.~Gasperini \& Veneziano 1993a, 1993b). The
application of the scalar fields to astrophysical objects is different
but the motivation for that can certainly be put along the lines of the
high energy and particle physics.

Although proposed thirty years ago, boson stars are so far only
hypothetical  objects (see the review papers by Jetzer 1992, Lee \&
Pang 1992, Liddle \& Madsen 1992, Mielke \& Schunck 1998). Apparently,
it took astronomers over  thirty years to actually detect neutron
stars since they were first proposed (see Shapiro \& Teukolsky 1983
for some historical notes). The first numerical studies of the boson
stars  were completed by Kaup (1968) and Ruffini \& Bonazzola
(1969). They found that the
mass of boson stars is of the order of $M_{Pl}^2/m$ where $M_{Pl}$
is the Planck mass. If self-interaction of the scalar particles is
assumed then boson stars have masses of the order of $M_{Pl}^3/m^2$ which
is approximately Chandrasekhar's mass provided the scalar field mass
$m$ is about 1GeV (Colpi et al. 1986). It suggests that boson stars
could be an important part of the dark matter. It was shown that {\em stable}
boson stars exist up to a maximal mass, quite similar to the neutron star
case (Kusmartsev et al. 1991) and the biggest mass star is called a
maximal boson star. They can be formed by gravitational
cooling, i.e., radiating away the surplus scalar matter
(Seidel \& Suen 1994).
Rotating boson stars change their form to become toroidal
(Schunck \& Mielke 1996, Ryan 1997, Yoshida \& Eriguchi 1997,
Schunck \& Mielke 1998).
Transparent boson stars were first studied by Schunck \& Liddle (1997).
It was proven that such boson stars both with and without self-interaction
could be very massive and proposed that boson stars could be an
alternative to black holes in the center of galaxies. Moreover, the radiation
from accretion disks within the gravitational potential of boson stars
would be gravitationally redshifted.

One of the main predictions of general relativity is the deflection of
light in gravitational fields of compact objects known as gravitational
lensing effect. Although it was first suggested in the context of general
relativity (Chwolson 1924, Einstein 1936), its approximate result for
the deflection of light by the Sun up to a constant factor {\it of two}
can also be derived within the framework of Newton's theory
(Schneider et al.~1992). Intensive studies of gravitationally lensed
objects began in the last decade or so and were mainly concentrated on the
weak gravitational lensing effect for which the angles of deflection are
very small up to a few arcseconds (for a review see Blanford \& Narayan
1992, Refsdal \& Surdej 1994, Narayan \& Bartelmann 1996). In general,
one can distinguish between lensing on non-cosmological and cosmological
distances. As for the former one can have lensing by a star
($M \approx 1 M_\odot$) in a galaxy with the deflection angles of the order
of milliarcseconds and by a galaxy ($M \approx 10^{11} M_\odot$) with the
deflection angles of the order of arcseconds.
As for the latter, model-dependent
luminosity distances has to be applied, but it gives the opportunity to
determine cosmological parameters such as the Hubble constant $H_0$,
the deceleration parameter $q_0$ and the cosmological constant $\Lambda$
(Turner et al.~1984, Gott \& Park 1989, Turner 1990, Fukugita et al.~1990).
A `giant' (cosmological) gravitational lens of the order of $157$ arcsecs has
also been detected but the immense deflection angle is hardly believed
to be the effect of lensing (Blanford et al.~1987).
In the case of weak lensing, the simple comparison with Newtonian theory
is valid, but all the Newton-based intuition fails for the cases of the
large deflection angles. If one has a black hole, or a strong compact object
such as a neutron star, then the gravitational lensing effects have to
be treated in fully relativistic terms. The deflection angles can be of
the order of many degrees allowing the light to orbit the lens many
times and multiple images can be formed (Cunningham \& Bardeen 1973,
Ftaclas et al.~1986, Ohanian 1987, Avni \& Shulami 1988, Nemiroff 1993,
D\c{a}browski \& Osarczuk 1995, Cramer 1997). These effects are
especially interesting for accretion disks (Luminet 1979, Bao et
al.~1994, Peitz \& Appl 1997).
Among some other lensing cases, it is also worth mentioning the
gravitational lensing by exotic matter such as cosmic strings
(Vilenkin 1984, Gott 1986) and by gravitational waves (Faraoni 1993).

The role of the scalar field in gravitational lensing has also been
investigated recently by Virbhadra et al.~(1998). They apply a real
massless scalar field into the Einstein equation and get a general static
spherically symmetric asymptotically flat solution with a naked
singularity which, apart from mass, has an extra parameter which they
called ``scalar charge''. They found a possibility of having four
images of a source at the observer position together with two Einstein
rings (a double Einstein ring).
In this paper, we investigate the gravitational lensing of a transparent
spherically symmetric boson star without self-interaction of the complex
scalar field. We can split the gravitational field into two regimes:
the inner part - where we have a smooth behavior of the curvature without
singularities or event horizon; the outer part - where the well-known
Schwarzschild solution applies. The boson star has no clearly defined
surface but an exponentially decreasing energy density. This allows to
define the radius of a star as corresponding to the radius of a sphere
which contains 99.9\% of the mass which gives almost no influence
of the inner parts of the boson star on the curvature of space outside.

The paper is organized as follows. In Section 2 we study geodesic motions
in the gravitational field of boson stars. In Section 3 we discuss the
effect of gravitational lensing by boson stars namely the number of
images of a point source and its magnification. In Section 4 we give
the discussion of the results. In the Appendix we discuss the form
of the lens equation both for small as well as for large deflection angles.
We propose new formulas for large deflection angle scattering of
light near compact lenses.

\section{Trajectories of Photons in Gravitational Field of a Boson Star}

Boson stars are astrophysical objects for which the source of gravity is a
massive complex scalar field $\Phi$ and which are the solutions of the
field equations  of Einstein gravity coupled to a scalar field given
by
\bea
R_{\mu \nu } - \frac{1}{2} g_{\mu \nu } R & = &
                  - 8\pi G T_{\mu \nu } (\Phi ) \; , \\
\Box \Phi + m^2 \Phi & = & 0 \; ,
\eea
where the energy-momentum tensor of matter composed of the scalar field is
\bea
T_{\mu \nu } = (\partial_\mu \Phi^\ast ) (\partial_\nu \Phi )
 - \frac{1}{2} g_{\mu \nu }
 \left [ g^{\sigma \rho } (\partial_\sigma \Phi^\ast )
  (\partial_\rho \Phi ) - m^2 |\Phi |^2 \right ].
\eea
In (1)-(3)
$
\Box = \partial_\mu
\Bigl [ \sqrt{\mid g \mid } \; g^{\mu \nu } \partial_\nu \Bigr ]/
\sqrt{\mid g \mid }
$
is the generally covariant d'Alembertian, $R_{\mu \nu }$ the Ricci
tensor, $R$ the curvature scalar, $g$ the determinant of the metric
$g_{\mu \nu }$, $m$ is the mass of the field $\Phi$ ($\hbar=c=1$),
and $G$ the Newtonian gravitational constant.
Stable states of the boson star exist (e.g. Kusmartsev et al.~1991).

For a static spherically symmetric boson star the metric is
\be
ds^2 = e^{\nu (r)} dt^2 - e^{\mu (r)} dr^2
  - r^2 ( d\theta^2 + \sin^2\theta \, d\varphi^2) \; .
\ee
The most general scalar field ansatz consistent with the static metric (4) is
\be
\Phi (r,t) = P(r) e^{-i \omega t} \; ,
\ee
where $\omega$ - the frequency of the scalar field. Notice that to find
finite mass solutions, the scalar field must carry a time-dependence of
the form (5), which leaves the energy-momentum tensor, and hence the metric,
time-independent. The solution of Virbhadra et al.~(1998) can be obtained
from equations (1)-(5) after putting $m = \omega = 0$ which makes the
scalar field $\Phi$ time-independent. The difference between both
solutions is that boson stars are non-singular solutions while Virbhadra's
et al.~solution possesses a naked singularity. Because of that boson star
solutions have much more in common with Schwarzschild than Virbhadra's
et al.~and the same should refer to any relativistic effects
under consideration.

The full set of Einstein equations (1) for the metric (4) reads
\bea
\nu' + \mu' & = & 8\pi G (\rho + p_r) r e^\mu
\; , \label{nula}\\
\mu' & = &  8\pi G \rho r e^\mu - \frac {1}{r}
(e^\mu - 1)
\; , \label{la}
\eea
the two further components giving degenerate equations because of
the Bianchi identities.

The differential equation for the scalar field with the ansatz (5) is
\be
P'' + \left ( \frac {\nu' - \mu'}{2} + \frac {2}{r} \right )
 P' + e^{\mu - \nu } \omega^2 P
- e^{\mu } m^2 P = 0  \; , \label{2ska}
\ee
and the non-vanishing components of the energy-momentum tensor (3) are
\bea
T_0{}^0 = \rho &=& \frac{1}{2} [ \omega^2  P^2(r) e^{-\nu }
   + P'^2(r) e^{-\mu } + m^2 P^2 ] \; , \\
T_1{}^1 = p_r &=& \frac{1}{2} [ \omega^2  P^2(r) e^{-\nu }
   + P'^2(r) e^{-\mu } - m^2 P^2 ] \; , \\
-T_2{}^2 = -T_3{}^3 = - p_\bot &=&
 \frac{1}{2} [ \omega^2  P^2(r) e^{-\nu }
   - P'^2(r) e^{-\mu } - m^2 P^2 ]
\eea
where $'=d/dr$. Note that the pressure is anisotropic; there are two
equations of state
$p_{r} = \rho - m^2 P^2$ and $p_\bot = \rho - m^2 P^2 - P'^2(r) e^{-\mu }$.

For the rest of our paper, we employ the dimensionless quantities $x=m
r$, $\sigma = \sqrt{4\pi G} \; P$. In order to obtain solutions which are
regular at the origin, we must impose the boundary conditions $\sigma
'(0)=0$ and $\mu (0)=0$.

At the distance $r$ one has
\be
e^{-\mu(r)} = 1 - \frac{2GM(r)}{r} \; , \quad
M(r) := \int\limits_0^r \rho y^2 dy
\ee
and far away from the star the metric is that of Schwarzschild with
\be
e^{\nu(r)}=e^{-\mu(r)} = 1 - \frac{2 G M}{r} \; .
\ee

\begin{figure}[t]
\centering
\leavevmode\epsfysize=14cm \epsfbox{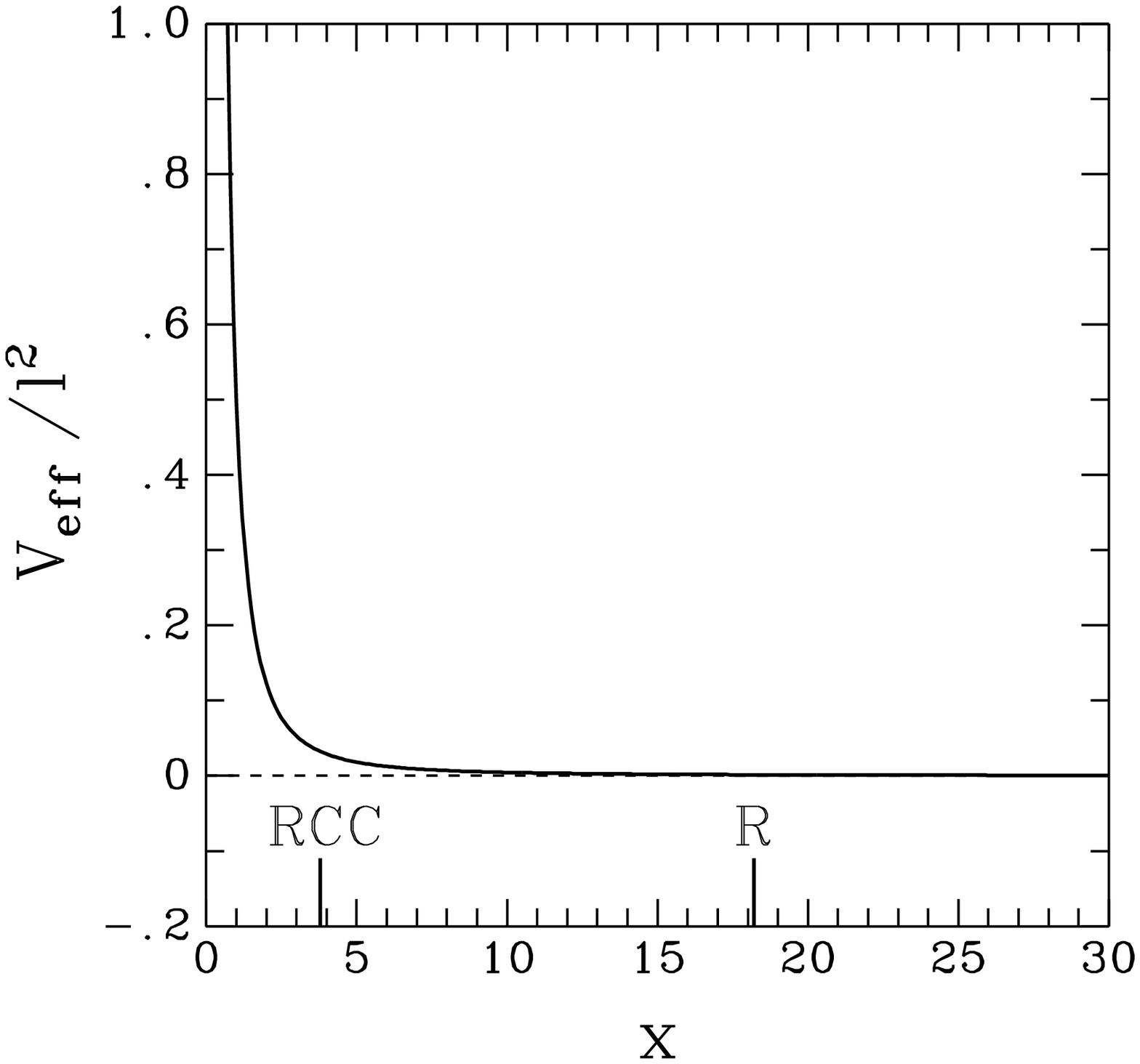}\\
\caption[]
{Effective potentials for null geodesics for
$\sigma (0)=\sqrt{4\pi P(0)}=0.1$.
The radius R of this boson star is at $x=18.2$ (99.9\% of the total mass).
$V_{\mbox{eff}}$ is measured in units of [$l^2$] and $x=mr$.
The position of the radial critical curve RCC is given (cf.~Table 2).
The radial caustic is at a large distance from the boson star, and is
not marked on this figure.}
\label{fig1}
\end{figure}

The main point of our analysis is the gravitational lensing properties
by a boson star and this requires the discussion of trajectories of photons
along the null geodesics described by null geodesic equations, i.e.,
\bea
\frac{1}{2} \left(\frac{dr}{d\tau}\right)^2 + V_{eff} & = &
  \frac{1}{2} F^2 e^{-\mu -\nu }
\label{null1} \; , \\
e^\nu \frac{dt}{d\tau} & = & F \; , \label{null2}\\
r^2 \frac{d\phi}{d\tau} & = & l \; , \label{null3}
\eea
where $F$ is the total energy and $l$ the angular momentum of the photon
(Wald 1984), $\tau$ is an affine parameter along null
geodesics and the effective potential in (\ref{null1}) is
\be
V_{eff} = e^{-\mu } \frac{l^2}{2 r^2} \; , \label{veff}
\ee
where $e^{\mu }$ is the globally regular metric component of the
boson star.
As mentioned before, outside the boson star one gets the well-known
Schwarzschild geometry and the formulas (\ref{null1})-(\ref{null3})
reduce to the standard ones as given, for instance, by
Shapiro \& Teukolsky (1983). The shape of the effective potential (17)
is given in Figure \ref{fig1}. Outside the boson star the potential is
of the Schwarzschild form $V_{eff} = l^2 (1-2M/x)/(2x^2)$
(e.g. Shapiro \& Teukolsky 1983). Inside, the effective potential has
a singularity at the center (for $l \neq 0$), i.e., no photon can be
captured by the boson star. Radial motion of a photon for $l=0$ is
insensitive to the potential as in the Schwarzschild case.

{}From now on, we assume that the light travelling from a source is
deflected by a boson star and the deflection angle is given by (Wald
1984, see also Fig. \ref{fig6})
\be
\hat{\alpha}(r_0) = \Delta \varphi(r_0) - \pi \; ,
\ee
where
\be
\Delta \varphi(r_0) = 2 \int\limits_{r_0}^\infty
\frac{e^{\mu /2}}{\sqrt{\frac{r^4}{b^2} e^{-\nu} - r^2}} dr \; ,
\ee
with the impact parameter
\be
b=l/F=r_0\exp{[-\nu(r_0)/2]} \; ,
\ee
and $r_0$ being the closest distance between a light ray and the center
of the boson star where we have
$V_{eff}(r_0) = (F^2/2) \exp[-\nu(r_0) - \mu(r_0)]$.

\section{Gravitational Lensing by a Boson Star}

The lens equation {\it for small deflection angles} may be expressed as
(e.g.~Narayan \& Bartelmann 1996, Virbhadra et al.~1998, see also the Appendix of this paper)
\be
\sin\lt(\vartheta-\beta\rt) = \frac{D_{ls}}{D_{os}} \sin\hat{\alpha} \, ,
\ee
where $D_{ls}$ and $D_{os}$ are distances from the lens (deflector)  to the
source and from the observer to the source, respectively,  $\beta$ denotes
the true angular position
of the source, whereas $\theta$ stands for the image positions.
One usually defines the reduced deflection angle to be
\be
\alpha \equiv \vartheta - \beta = \sin^{-1}\lt(\frac{D_{ls}}{D_{os}}
\sin\hat{\alpha}\rt) \, .
\ee
However, the equation (21) relies on substitution of the distance from the
source to the point of minimal approach by the distance from the lens to the
source $D_{ls}$(see Narayan
\& Bartelmann 1996). {\it For large deflection angles} the distance $D_{ls}$
cannot be considered a constant but it is a function of the deflection angle
so that the form of the lens equation changes into (see Appendix)
\begin{equation}
\sin{\alpha} = \frac{D_{ls}}{D_{os}} \cos{\vartheta}
\cos \lt[ \mbox{arsin}
\lt ( \frac{D_{os}}{D_{ls}} \sin (\vartheta - \alpha ) \rt) \rt]
\lt[\tan{\vartheta} + \tan(\hat{\alpha} - \vartheta)
\rt] \, .
\end{equation}
The reduced deflection angle can be kept defined as in (22), i.e.,
$\alpha \equiv \vartheta - \beta$.

The magnification of images is given by (Narayan \& Bartelmann 1996)
\be
\mu = \lt( \frac{\sin{\beta}}{\sin{\vartheta}} \
\frac{d\beta}{d\vartheta} \rt)^{-1}.
\ee
The tangential and radial critical curves (TCC and RCC, respectively)
follow from the singularities of the tangential
\be
\mu_t \equiv \lt(\frac{\sin{\beta}}{\sin{\vartheta}}\rt)^{-1} ,
\ee
and the radial magnification
\be
\mu_r \equiv \lt(\frac{d\beta}{d\vartheta}\rt)^{-1}.
\ee
{}From the geometry of the lens we have the relation (cf. Fig.
\ref{fig6})
\be
\sin{\vartheta} = \frac{b}{D_{ol}} = \frac{r_0}{D_{ol}} \exp{[-\nu(r_0)/2]} .
\ee
In order to obtain magnification (23) the first derivative of the deflection
angle (18) with respect to $\vartheta$ has to be calculated and it is
given by
\be
\mu_r^{-1} = \frac{d\beta}{d\vartheta} = 1 -
\frac{d\alpha(\vartheta)}{d\vartheta} = 1-\frac{D_{ls}}{D_{os}}
 \frac{\cos \hat \alpha }{\cos \alpha }
\frac{d\hat \alpha}{dr_0} \frac{dr_0}{d\vartheta} \;  ,
\ee
where
\be
\frac{dr_0}{d\vartheta} = \frac{1}{2} \exp{[-\nu(r_0)/2]}
\frac{2 - r_0 \frac{d\nu(r_0)}{dr_0}}
{\sqrt{1 - \frac{r_0^2}{ D_ol^2} \exp{[-\nu(r_0)]} } }  \;   ,
\ee
and $d\hat{\alpha}/dr_0$ can be calculated by parametric differentiation of
equation (19) with respect to $r_0$.

\begin{figure}[t]
\centering
\leavevmode\epsfysize=14cm \epsfbox{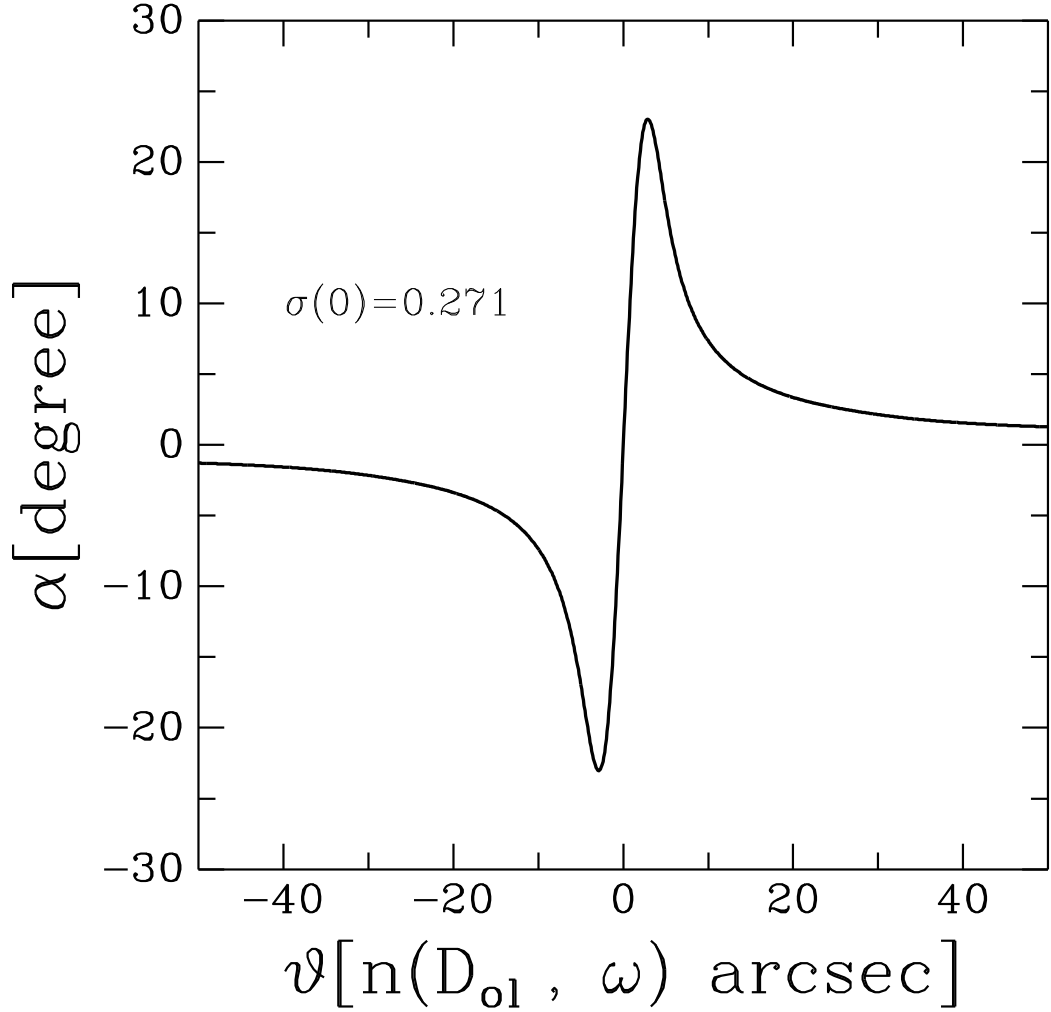}\\
\caption[]
{Reduced deflection angle for a maximal boson star.
The angle $\vartheta$ is defined in arcsec
times the function $n$ which depends on the observer-to-lens distance
$D_{ol}$  and the scalar field frequency. We have chosen
$D_{ls}/D_{os}=1/2$ in Eq.~(23) or $D_{ps}/D_{os}=1/2$ in Eq.~(A19),
respectively}.
\label{fig2}
\end{figure}

We performed numerical calculations of the reduced deflection angle
(22) and presented the results in Figure \ref{fig2} and Table
\ref{table1} and \ref{table2}. We have assumed that a boson star
(lens) is half-way between the observer and the source, hence
$D_{ls}/D_{os}=1/2$ in equations (22) and (27).

The plot of the reduced deflection angle $\alpha = \alpha(\vartheta)$
for the  maximal boson star with the initial value of $\sigma
(0)=0.271$ is given in Figure \ref{fig2}. The maximal reduced
deflection angle is presented in Table \ref{table1}. The biggest
possible value of the deflection angle is $23.03$   degrees with an
image at about $\vartheta=n\cdot2.8878$ arcsecs where $n =
n(D_{ol},\omega)$ is a function of the distance from the observer to
the lens $D_{ol}$ and the value of the scalar field frequency $\omega$
(the latter refers to the size of the boson star). Such big deflection
angles are possible since boson stars are very strong relativistic
objects.

This is especially clear after analysing Figs.~4a and 4b. In
Fig.~4a we show the mass distribution $M = M(x)$ (cf.~Eq.(12)) of
a boson star (which is an extended object, like a sun) of maximal mass,
the so-called Kaup limit $0.633 M_{pl}^2/m$. In Fig.~4b we present
the plot of the `coefficient of relativisticity' $2M/x$ of a boson star
(solid line) against a black hole (dashed line). It demonstrates
that such a boson star has a gravitational field whose strength is
comparable to the strength of the gravitational field of a black hole.

The light passing through the interior of a star is
effected if $\vartheta \leq 20 n(D_{ol},\omega )$ arcsecs while for
$\vartheta > 20 n(D_{ol},\omega )$ arcsecs the effect is almost the
same as in Schwarzschild spacetime (with increasing
distance to a boson star the difference with Schwarzschild spacetime
is negligible). We have numerically and analytically checked that far away from
a boson star one has the deflection angle to be equal to $2R_{g}/b$,
where $R_g$ is the Schwarzschild radius.

\begin{figure}[t]
\centering
\leavevmode\epsfysize=7cm \epsfbox{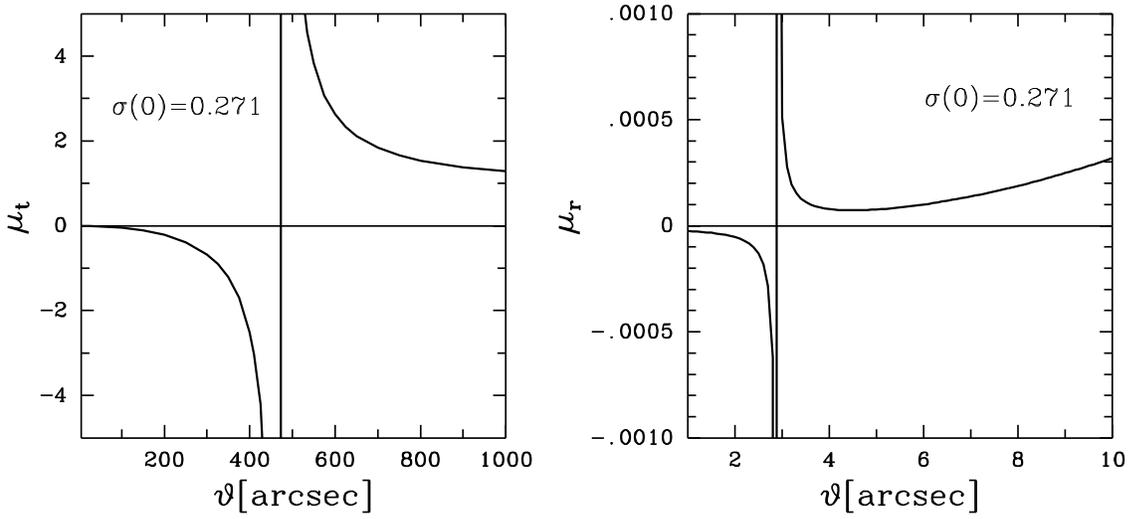}\\
\caption[]
{Tangential (left) and radial (right) magnification for the case of a
maximal boson star with $\sigma(0)=0.271$. The distance was chosen in such a
way that the factor $n$ equals 1 and, hence, $\vartheta$ is measured
in arcsec. The Einstein ring (TCC) appears at 472.5 arcsecs and the radial
critical curve (RCC) at 2.8877 arcsecs (cf.~Table 2).}
\label{fig3}
\end{figure}

The tangential and radial magnification $\mu_t$ and $\mu_r$,
respectively, are given in Figure \ref{fig3}. In order to make a plot
the distance to the boson star has to be fixed. In Figure \ref{fig3}
we have chosen the distance factor $n=1$ so that $\vartheta $ is
measured in arcsecs. Then, the Einstein ring (TCC) is found at about
$472.5$ arcsecs and the radial critical curve (RCC) at about $2.8877$
arcsecs (cf.~Table \ref{table2}).
If the distance factor $n=60$, for example, the star is closer to the
observer and so $\vartheta$ is measured in arcmins. Then,
we find that the TCC is at $61.73$ arcmins and the RCC is at $2.8844$ arcmins
as in the second line of Table \ref{table2}. Further
examples for different boson stars including Newtonian ones, with
small initial value of the scalar field $\sigma $ can be extracted
from Table \ref{table2}. It is remarkable to notice that for a fixed
distance the Einstein rings (TCCs) decrease with decreasing initial
values of central densities of the boson star and the RCCs increase
until both vanish.

It is useful to notice that the radial critical curve (RCC) appears within
the radius of a boson star. This is obvious since RCCs are in general possible
only for transparent astronomical objects. It can easily be seen from the
plot of $V_{eff}(x)$ in Fig.~1 and from the plot of $M(x)$ in Fig.~4a.

Using the standard notation (e.g.~Narayan and Bartelmann 1996) we should call
TCCs and RCCs the critical lines (they both are circles of angular size
$\vartheta$ in the lens plane given in Table 2) while the caustics are the lines of the angular size
$\beta$ (source angle) in the source plane corresponding to these TCCs
and RCCs. In the case of a transparent boson star, the caustic
corresponding to a TCC is a point, while the caustic corresponding to
a RCC is a circle of angular size $\beta$ (see Figure \ref{fig5}).

Finally, let us discuss the characteristics of the boson stars
producing the  deflection angles under consideration. Both the
observer-to-lens-distance $D_{ol}$ and the mass of the boson star play
a role therein. If $\vartheta$ is chosen to be of the order of arcsecs
(distance factor $n=1$), then the distance
$D_{ol}$ is measured in units of 206265$/\omega$.
Under the assumption that the mass of the
boson star is 10$^{10}$M$_{\odot}$ one has $\omega \sim 10^{-15}$cm.
Then distance $D_{ol}$ is about 100pc.
If the distance factor is $n=10^{-3}$, and so
$\vartheta $ measured in milliarcsecs, the boson-star-lens is at about
100kpc.

\section{Summary}

In this paper we discussed the possibility for a boson star to be a
gravitational lens. We assumed a boson star to be transparent and
calculated the deflection angles of a point source for various ranges of
the parameters of the models. All qualitative features of a non-singular
circularly symmetric transparent lens can be revealed. We derived the
lens equation and plotted the lens curve for a maximal boson
star. We have shown that there are typically three images, two of them
being inside the Einstein
radius and one outside. In most of the cases (depending on the parameters
of the models) there is an Einstein ring with infinite tangential
magnification (tangential critical curve - TCC) and the angle for which two
internal images merge leading to infinite radial magnification
(radial critical curve - RCC). The existence of the radial critical curve
is the result of the transparency of a boson star. The deflection angles caused by boson stars
can be very large (of the order of degrees) which allows to compare
observationally these lenses with other known lenses. Also, the
appearance of the radial critical curve would differ boson stars from
other extended and non-transparent lenses. For a black hole or a
neutron star, the radial critical curve does not exist because it is
inside the event horizon of the star. The assumption of transparency
for the boson star is essential
if one wants to find two bright images near the center of a boson
star and the third image at some very large distance from the
center. This happens in relativistic case. In the example of a
maximal boson star case ($\vartheta $ measured in arcsecs), the source
would be at an angular distance of slightly above $23$ degrees from the lens
and the two images inside the boson star at about $2.8$ arcsecs. An
interesting point can be made if one considers an extended source.
In such a case one finds the two radially and tangentially elongated
images very close to each other. Then, looking along the line defined
by these two images one could find the third one with very large
deflection angle and this would be the case for a boson star.
However, this third image would be highly demagnified as shown in
our Table \ref{table3}.

Another comment refers to the possibility for the scalar field
to have an additional $U(1)$ symmetry, which yields the electromagnetic
field to be present inside a star. This possibility is simply a charged boson
star which has already been investigated by Jetzer \& van der Bij (1989). It
has also been shown
that stable charged boson stars exist. However, the charged scalar
particles interact with photons passing
through the star and such boson stars become opaque. In analogy to what happens
for a neutral boson star, it is clear that the exterior of a charged boson
star can be described by the Reissner-Nordstr\"om geometry, and the
deflection angles are qualitatively the same as for a charged neutron
star - the case already considered in the literature (D\c{a}browski
\& Osarczuk 1995). It is obvious that for a charged (so opaque)
boson star, only two images are possible.

Finally, we hope that one of the proposed possibilities will be useful
for some observational investigations. We also emphasize that the
possible detection of boson stars would make a very strong evidence
for scalar fields in the universe supporting most of the very early
universe scenarios.

\acknowledgements

MPD acknowledges the support of NATO/Royal Society and the hospitality of the
Astronomy Centre at the University of Sussex. FES was supported by a
Marie Curie research fellowship (European Union TMR programme). We
would like to thank Andrew Barber, John Barrow, Andrew Liddle,
Eckehard Mielke, Sophie Pireaux, Jerzy Stelmach, Kandaswamy
Subramanian, and Diego Torres for helpful discussions and comments.
We also thank the anonymous referee for raising the important point about
the form of the lens equation for large deflection angles. MPD
wishes to thank once more Kandaswamy Subramanian for the discussion
about this particular point during the workshop at Isaac Newton Institute,
Cambridge, in the summer 1999.

\appendix
\vspace{0.6cm}
\section{Lens Equation for Large Deflection Angles}
\vspace{0.6cm}

In Figure 6, $D_{ol}$ is the distance from the observer to the
lens, $D_{ls}$ is the distance from the lens to the source and $D_{os}$
is the distance from the observer to the source. The angle $\beta$
denotes the true angular position of the source whereas $\vartheta$
measures the angle of the source position. By construction, the
source S is always be within a plane with constant distance to the
observer. Note that for small
deflection angles the distance $D_{ps}$ from the source to the point
of closest approach can be replaced by $D_{ls}$ and so according
to the sine formula one is able to get the lens equation in the
form presented in standard literature which is Equation (21) of
Section 3.

This derivation, however, mainly relies on the fact that for small
deflection angles the distances $D_{ol}, D_{os}$ and $D_{ls}$ can
be considered as the distances along the line of sight of the
observer and so $D_{os} \approx D_{ol} + D_{ls}$. For large
deflection angles the last relation is not true and both $D_{ls}$
and $D_{os}$ depend essentially on the deflection angle
$\hat{\alpha}$, i.e.,
\begin{equation}
D_{os} < D_{ol} + D_{ls}  \, .
\end{equation}
Then, for large deflection angles equation (21) has to be
replaced by a generalized formula. An example of such a formula
has recently been proposed by Virbhadra \& Ellis (1999), however,
their definition of distances does not fulfill relation (A1)
and so we propose another formula which relies on the correct
physical definition of distances.

For the derivation of the lens equation the following angles
should be calculated: $\psi, \omega$ and $\sigma$. These angles
are presented in Fig.~6. Simple calculation shows that
$\omega = \hat{\alpha} - \vartheta - \beta$,
$\psi = \pi/2 + \vartheta - \hat{\alpha}$ and
$\sigma = \hat{\alpha} - \vartheta$.
It can be shown that the azimuthal deflection angle is equal to
\begin{equation}
\delta{\varphi} = \lt(\pi/2 - \varepsilon\rt) + \hat{\alpha} +
\lt(\pi/2 - \vartheta\rt)
\end{equation}
so that the deflection angle is (Ohanian 1987)
\begin{equation}
\hat{\alpha} = \delta\varphi - \pi + \varepsilon + \vartheta =
\bar{\alpha} + \varepsilon + \vartheta \, ,
\end{equation}
where $\bar{\alpha}$ is the deflection angle SLM as defined in Fig.~6 and
$\Delta \varphi \approx \delta \varphi$.
However, from the geometry of the bending effect we have,
\begin{equation}
\sin{\vartheta} = \frac{b}{D_{ol}} , \hspace{1.0cm} \sin{\varepsilon} =
\frac{b}{D_{ls}}  \, .
\end{equation}
In real situations the distances $D_{ol}$ and $D_{ls}$ are
much bigger than the impact parameter $b$ and so the trigonometric
functions of angles in (A4) can be approximated by angles
themselves and the relation (A3) reads
\begin{equation}
\hat{\alpha} = \bar{\alpha} + \frac{b}{D_{ls}} +
\frac{b}{D_{ol}} \, ,
\end{equation}
so in practice one can replace the deflection angle $\bar{\alpha}$
by the deflection angle $\hat{\alpha}$ defined by Eq.~(18) and vice versa.
{}From Fig.~6 we can write down the following relations
\begin{equation}
\cot{\psi} = \frac{SK}{KP} ,\hspace{1.cm} \cot{(\pi/2 - \vartheta)} =
\frac{KI}{KP}  \, ,
\end{equation}
so that
\begin{equation}
SI = SK + KI = KP \lt(\cot{\psi} + \cot{\gamma}\rt) \, .
\end{equation}
On the other hand
\begin{equation}
\tan{\beta} = \frac{MS}{MO} \, , \hspace{1.cm} \tan{\vartheta} = \frac{IS
- MS}{MO} \, ,
\end{equation}
so that
\begin{equation}
\tan{\beta} = \frac{IS}{MO} - \tan{\vartheta} \, .
\end{equation}
Note that ($KP \approx ML$)
\begin{equation}
KP = D_{ls}\cos{\bar{\alpha}} \, , \hspace{1.cm} MO = D_{os}
\cos{\beta}  \, ,
\end{equation}
and so
\begin{equation}
SI = D_{ls} \cos{\bar{\alpha}} \lt( \cot{\psi} + \cot{\gamma}
\rt) \, .
\end{equation}
Since
\begin{equation}
\cot{\psi} = \tan{(\hat{\alpha} - \vartheta)} \, ,
\end{equation}
and
\begin{equation}
\cot{\gamma} = \tan{\vartheta} \, ,
\end{equation}
then we have from (A9)
\begin{equation}
\tan{\beta} = \frac{D_{ls}\cos{\bar{\alpha}}}{D_{os}\cos{\beta}}
\lt[ \tan{\vartheta} + \tan(\hat{\alpha } - \vartheta) \rt] -
\tan{\vartheta} \, .
\end{equation}
Note that here $\beta$ ($\alpha \equiv \vartheta + \beta$ - cf.~Fig.~6)
should be taken to be
negative in order to get the relation to the formula of Virbhadra and Ellis
(1999). After using trigonometric relations one can reduce (A14) to the form
\begin{equation}
\sin{\alpha} = \frac{D_{ls}}{D_{os}} \cos{\vartheta}
\cos{\bar{\alpha}} \lt[\tan{\vartheta} + \tan(\hat{\alpha} - \vartheta)
\rt] \, ,
\end{equation}
or
\begin{equation}
\sin{\alpha} = \frac{D_{ls}}{D_{os}}
\frac{\cos{\bar{\alpha}}}{\cos{\vartheta}}
\frac{\tan \hat{\alpha}} {\lt( 1 +
\tan{\hat{\alpha}}\tan{\vartheta} \rt)} \, .
\end{equation}
The equation (A16) is the lens equation for large deflection angles and
it can be reduced to the equation (21) provided the angle $\vartheta$ is small.
Using the sine formula from the triangle (OSL) we have
\begin{equation}
\sin \bar \alpha = \frac{D_{os}}{D_{ls}} \sin \beta  \, ,
\end{equation}
so that a reduced deflection angle $\alpha$ is determined by the nonlinear
formula
\begin{equation}
\sin{\alpha} = \frac{D_{ls}}{D_{os}} \cos{\vartheta}
\cos \lt[ \mbox{arsin}
\lt ( \frac{D_{os}}{D_{ls}} \sin (\vartheta - \alpha ) \rt) \rt]
\lt[\tan{\vartheta} + \tan(\hat{\alpha} - \vartheta)
\rt] \, .
\end{equation}

If the source S is not within a plane with constant distance to the observer,
a different lens equation can be derived. The situation is also presented in
Figure 6. From the two triangles (OSJ) and (PSJ), the lens equation reads
\begin{equation}
\sin\lt(\vartheta+\beta\rt) = \frac{D_{ps}}{D_{os}} \sin\hat{\alpha} \, ,
\end{equation}
where $D_{os}$ is the observer-source distance and
$D_{ps}$ the distance between source and point P, the closest approach
to the lens.

Fig.~2 describes both physically different situations represented by (A18) and
(A19) because the corrections in the reduced deflection angles are
in fact negligible (the correction appears in the third digit) due to the assumed large
source-lens and observer-lens distances.

%==================================================

%\pagebreak

\begin{figure}[t]
\centering
\leavevmode\epsfysize=14cm \epsfbox{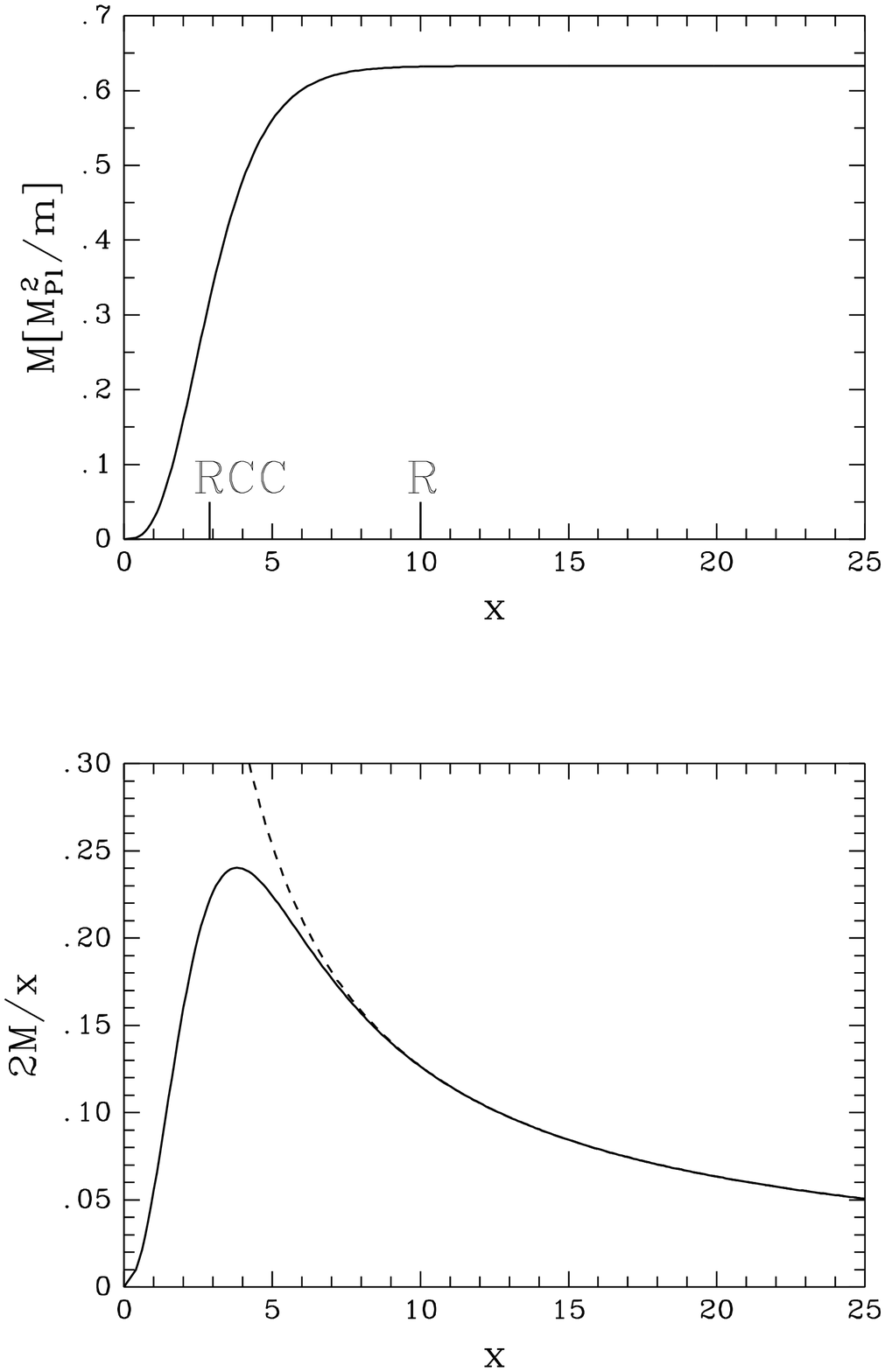}
\caption[]\\
a) The mass distribution $M(x)$ ($x = mr$) for a maximal boson star.
The mass becomes constant at the Kaup limit $0.633M_{pl}^2/m$, which
happens close to the radius R.\\
b)
The `coefficient of relativisticity' $2M/x$ of a boson star (ratio of the
Schwarzschild radius to the radius determined for a maximal boson star
--- solid line)
and of a black hole with the same mass (dashed line). The dashed line
decreases from the horizon of the black hole ($2M/x = 1$) and exhibits
the Keplerian fall for a constant mass distribution.
\label{fig4}
\end{figure}

\begin{figure}[t]
\centering
\leavevmode\epsfysize=7cm \epsfbox{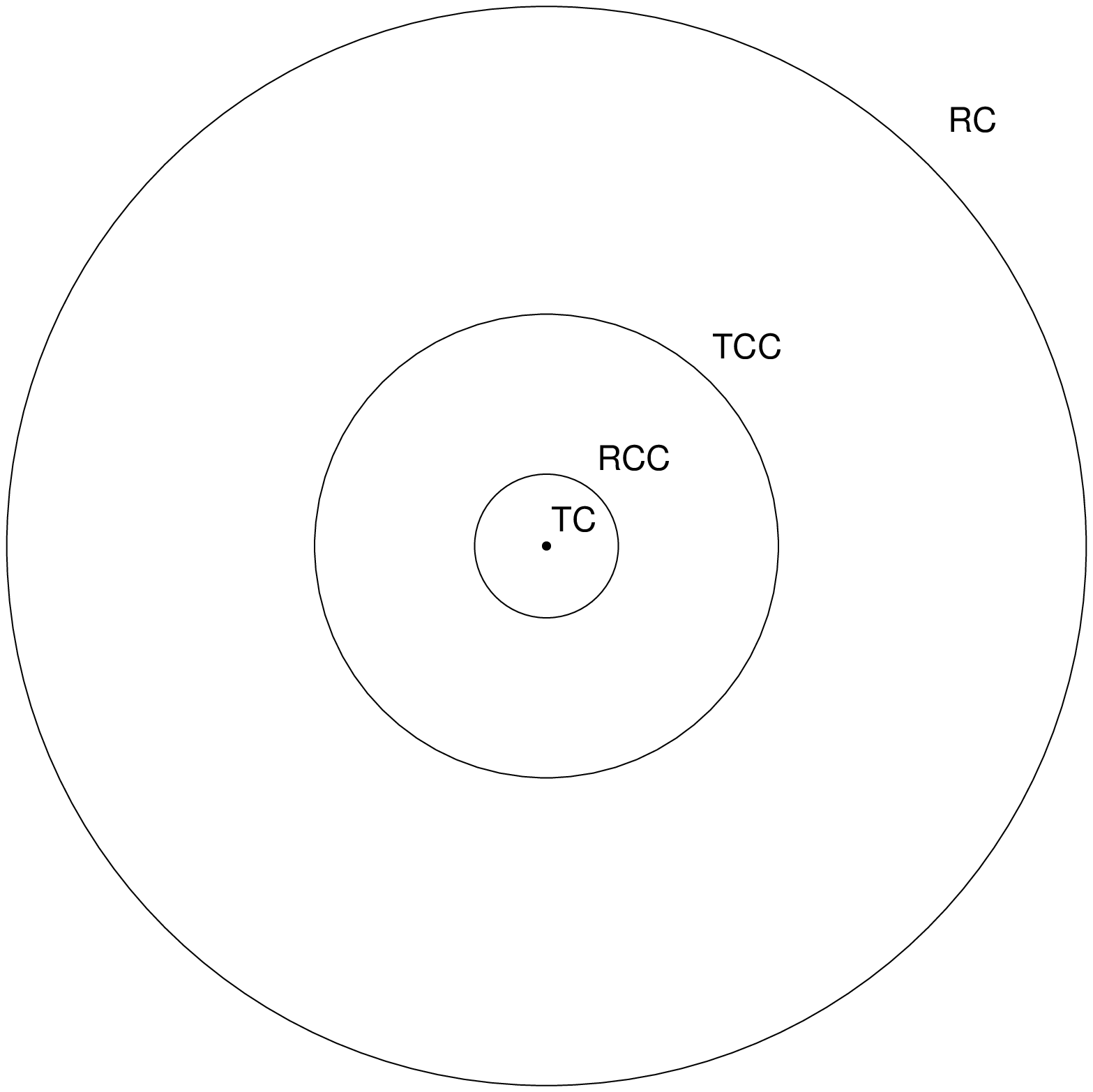}\\
\caption[]
{Critical lines (CC) and caustics (C) for maximal boson star ---
explicit values are taken from first line of Table 2.}
\label{fig5}
\end{figure}

\begin{figure}[t]
\centering
\leavevmode\epsfysize=12cm \epsfbox{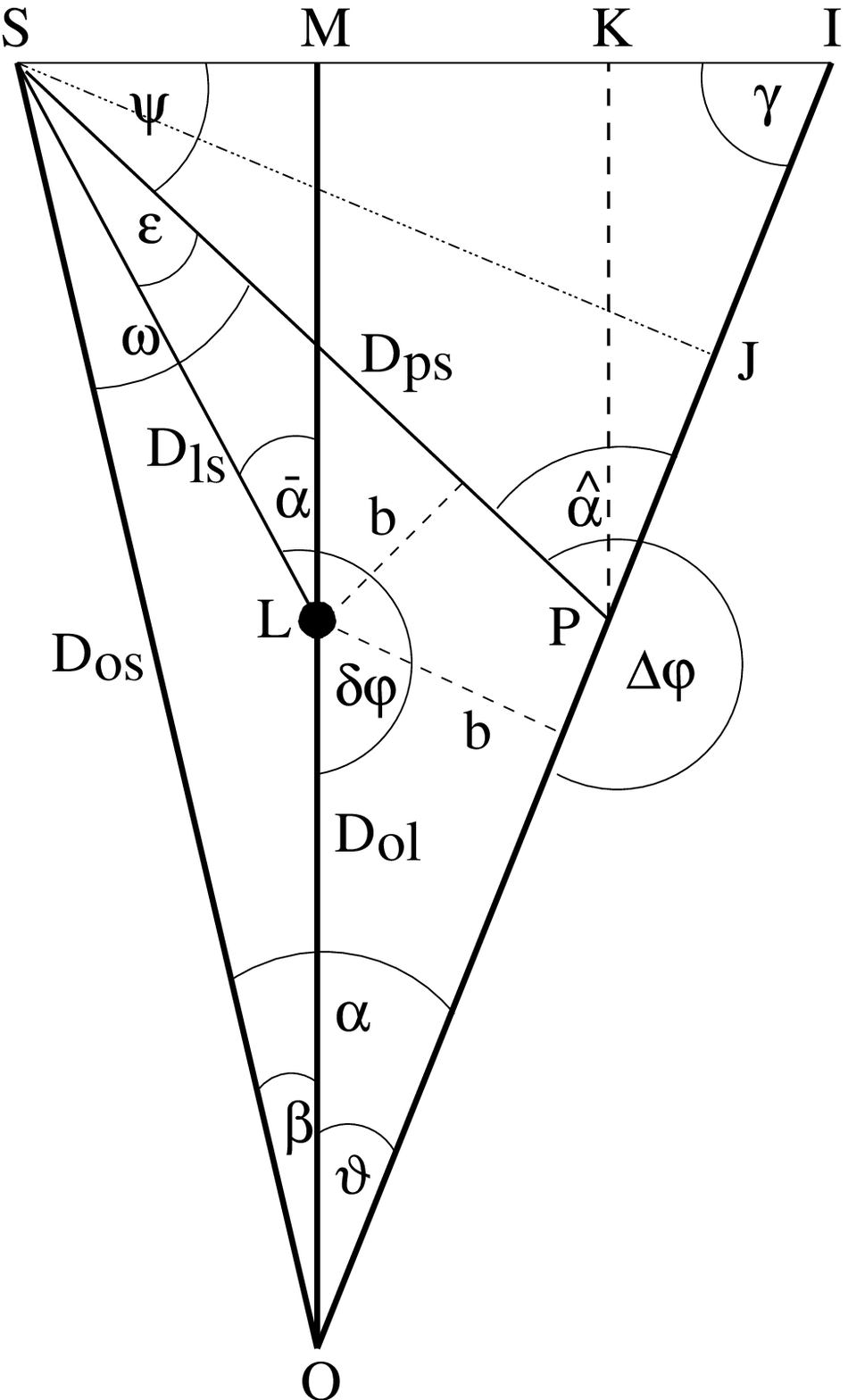}\\
\caption[]
{Schematic lens diagram as described in the text of Appendix.}
\label{fig6}
\end{figure}

\def\rrr{\rule[-3mm]{0mm}{8mm}}
\begin{table} [tbh]
\begin{center}
\begin{tabular}{lrr} \hline\hline
\rrr $\sigma(0)$ & $\alpha_{max}$ & $\vartheta /n(D_{ol},\omega )$ \\ \hline
\rrr   0.271     &   23.0309$^{o}$  &   2.8878''  \\
\rrr   0.1       &    8.6944$^{o}$  &   5.4523''  \\
\rrr   10$^{-3}$ &    5.0786'       &  59.4811''  \\
\rrr   10$^{-5}$ &    3.0183''      & 591.2845''  \\
\hline\hline
\end{tabular}
\end{center}
\begin{center}
\parbox {13cm} {
\caption {}\label{table1}
{Maximal reduced deflection angle $\alpha_{max}$ and the positions
$\vartheta $ for different boson stars. The deflection for Newtonian
boson stars is of the order of arcsecs while for general relativistic
boson stars it is of the order of degrees. The position of a maximum
is always well inside boson star as can be concluded from the values
of its radius (99.9\% of the total mass):
$x=$10.0, 18.2, 194.5, 1611.8 for $\sigma (0)=$0.271,
0.1, 10$^{-3}$, 10$^{-5}$. }}
\end{center}
\end{table}

\vfill
\pagebreak

\def\rrr{\rule[-3mm]{0mm}{8mm}}
\begin{table} [tbh]
\begin{center}
\begin{tabular}{lrrcr} \hline\hline
\rrr                   & $\mu_t$ & $\mu_r$& [$\vartheta $] &
     radial caustic [$\beta$]\\ \hline
\rrr $\sigma(0)=0.271$ &   8.695 & 2.696  & degree & 20.2399$^{o}$ \\
\rrr                   &   61.73 & 2.8844 & arcmin & 22.9825$^{o}$ \\
\rrr                   &  472.5  & 2.8877 & arcsec & 23.0298$^{o}$ \\
\hline $\sigma(0)=0.1$ &   7.795 & 3.7902 & degree & 4.1241$^{o}$ \\
\rrr                   &   59.30 & 5.4171 & arcmin & 8.6039$^{o}$ \\
\rrr                   &  454.25 & 5.4516 & arcsec & 8.6929$^{o}$ \\
\hline
\rrr $\sigma(0)=10^{-3}$ &    none &   none & degree & none \\
\rrr                     &    none &   none & arcmin & none \\
\rrr                     &  163.70 & 52.058 & arcsec & 4.1518' \\
\hline
\rrr $\sigma(0)=10^{-5}$ &    none &   none & arcsec,arcmin,degree &
     none \\
\hline\hline
\end{tabular}
\end{center}
\begin{center}
\parbox {14cm} {
\caption {}\label{table2}
{Positions of tangential and radial critical curves
(singularities of tangential and radial magnification, cf.~Figure \ref{fig3})
and values for radial caustics. The column for [$\vartheta $] gives
the units for the numbers of $\mu_t$ and $\mu_r$ in the same line.
The distance factor $n$ has the values 1, 60, 3600 for the units
arcsec, arcmin, degree, respectively. If the gravitational field of a
star is too weak, and/or the star is too close to the observer, there are no
critical curves at all. The numbers presented give always a lower limit
of the actual singularity value, namely one with higher negative $\mu_t$
or $\mu_r$.}}
\end{center}
\end{table}

\vfill

\def\rrr{\rule[-3mm]{0mm}{8mm}}
\begin{table} [tbh]
\begin{center}
\begin{tabular}{cr} \hline\hline
\rrr $\vartheta$    & $\mu = \mu_t \cdot \mu_r $ \\ \hline
\rrr   2.85''    &  $4.8 \times 10^{-8}$ \\
\rrr   2.88''    &  $2.0 \times 10^{-7}$ \\
\rrr   2.89''    & $-1.2 \times 10^{-5}$ \\
\rrr   2.90''    & $-1.9 \times 10^{-7}$ \\
\rrr   2.95''    & $-3.3 \times 10^{-8}$ \\
\hline\hline
\end{tabular}
\end{center}
\begin{center}
\parbox {13cm} {
\caption {}\label{table3}
{Magnification of a finite size source across a radial caustic (RC)
for the case of $\sigma (0)=0.271$, RC at about 23.0298$^o$,
and RCC at about 2.8877'' (cf.~Table \ref{table2}).
It clearly shows that the image is highly demagnified (with the exception of
the theoretically infinite value at RC).}}
\end{center}
\end{table}

\vfill

\end{document}